\documentclass[lettersize,journal]{IEEEtran}
\usepackage{amsmath,amsfonts}
\usepackage{algorithmic}
\usepackage{algorithm}
\usepackage{array}
\usepackage[caption=false,font=normalsize,labelfont=sf,textfont=sf]{subfig}
\usepackage{textcomp}
\usepackage{stfloats}
\usepackage{hyperref}
\usepackage{verbatim}
\usepackage{graphicx}
\usepackage{cite}
\usepackage{multirow}
\usepackage{makecell}
\usepackage{xcolor}
\usepackage[table]{xcolor}
\usepackage{bm}
\hypersetup{colorlinks=true,linkcolor=blue,urlcolor=blue}

\begin{document}

\title{Harmonic-Percussive Disentangled Neural \\ Audio Codec for Bandwidth Extension}

\author{Benoît Giniès, Xiaoyu Bie, Olivier Fercoq \& Gaël Richard \IEEEmembership{Fellow, IEEE}\\
LTCI, Télécom Paris, Institut Polytechnique de Paris \\
Palaiseau, France \\
\texttt{benoit.ginies@telecom-paris.fr}
}

\maketitle

\begin{abstract}
Bandwidth extension, the task of reconstructing the high-frequency components of an audio signal from its low-passed counterpart, is a long-standing problem in audio processing.
In this work, we extend recent advances in neural architectures by framing bandwidth extension as an audio token prediction problem. Specifically, we train a transformer-based language model on the discrete representations produced by a disentangled neural audio codec, where the disentanglement is guided by a Harmonic–Percussive decomposition of the input signals, highlighting spectral structures particularly relevant for bandwidth extension. Our approach introduces a novel codec design that explicitly accounts for the downstream token prediction task, enabling a more effective coupling between codec structure and transformer modeling. This joint design yields high-quality reconstructions of the original signal, as measured by both objective metrics and subjective evaluations. These results highlight the importance of aligning codec disentanglement and representation learning with the generative modeling stage, and demonstrate the potential of global, representation-aware design for advancing bandwidth extension.
\end{abstract}


\section{Introduction}
\IEEEPARstart{B}{andwidth} extension seeks to reconstruct the high-frequency content of a signal from its low-frequency representation. This problem can be viewed as a form of inpainting where missing information is recovered from degraded observations.
Applications arise in telecommunication, where it improves speech quality \cite{chennoukh2001speech}, as well as in music restoration \cite{moliner2022behm}. While early methods relied on handcrafted signal processing techniques \cite{Dietz2002SpectralBR}, recent approaches leverage neural networks \cite{pulakka2011bandwidth} that can learn efficient internal representations and achieve significantly better results.

Neural audio codecs, based on the VQ-VAE paradigm \cite{oja1982simplified,van2017neural}, provide low-bitrate latent representations while preserving high reconstruction quality. Beyond compression, the resulting latents have proven effective for downstream tasks such as audio generation \cite{borsos2023audiolm}. In parallel, advances in transformer architectures and large language models, combined with neural audio codecs, have established strong baselines in both speech \cite{wang2023neural} and music generation \cite{copet2023simple}.

Despite their effectiveness, codec-based representations often lack interpretability and versatility. This has motivated refined architectures that enforce semantically meaningful latents by reshaping training objectives and model design. Such disentangled representations, tailored to downstream tasks, can be exploited to improve task performance. Disentanglement has been explored in diverse audio processing settings, from specialized applications such as singing voice conversion \cite{takahashi2021hierarchical} to more general approaches \cite{hsu2023disentanglement}.

In this work, we leverage recent advances in language modeling over neural audio codec representations to address the bandwidth extension task. We developed a codec architecture specifically designed for bandwidth extension, one that learns a structured and spectrally meaningful latent representation. 
While prior studies have often treated codec structure as fixed and external to language model training, we propose integrating the codec design directly into the prediction pipeline. 

We first design the Harmonic-Percussive disentangled codec (HP-codec), a neural codec that explicitly separates high- and low-frequency components and further factors the latent space to capture harmonic and percussive structures, whose characteristic cross-band patterns improve the predictability of high-frequency components from their low-frequency counterparts. Building on this representation, we introduce HP-codecX, a bandwidth extension model based on a transformer language model whose architecture is adapted to HP-codec’s structure. The model is trained to predict high-frequency content from HP-codec’s low-frequency latents, thereby addressing the bandwidth extension task.

Our main contributions are as follows:
\textbf{(1)} We introduce HP-codec, a structure-informed disentangled neural audio codec that leverages an Harmonic–Percussive decomposition of audio signals.
\textbf{(2)} We adapt its latent representation to a language modeling task aligned with bandwidth extension.
\textbf{(3)} We design and train a multi-branch language model tailored for bandwidth extension.
\textbf{(4)} We demonstrate state-of-the-art bandwidth extension performance, achieving competitive results across both objective metrics and human listening tests. We distribute our implementation and pretrained weights on our \href{https://github.com/Harmonic-Percussive-Bandwidth-Extension/HP-codecX}{\underline{github}}\footnote{{https://github.com/Harmonic-Percussive-Bandwidth-Extension/HP-codecX}}, and audio examples on our \href{https://harmonic-percussive-bandwidth-extension.github.io/}{\underline{companion website}}\footnote{https://harmonic-percussive-bandwidth-extension.github.io/}.

\section{Related Work}
\subsection{Neural Audio Codecs and Disentanglement}

Recent representation learning paradigms integrate data-driven autoencoders\cite{kingma2013auto}, followed by residual vector quantization (RVQ) between encoder and decoder \cite{van2017neural}. Modern neural audio codecs typically combine an encoder, RVQ, and decoder, trained with composite objectives often including adversarial losses \cite{zeghidour2021soundstream, defossez2022high, kumar2023high}. While most of these models operate in the audio (waveform) domain, other work have focused on adapting this approach to the spectral domain \cite{li2025spectrostream}.

Subsequent research has refined codec architectures to address specific limitations. For example, a differentiable quantization mechanism to eliminate the stop-gradient trick was proposed in \cite{takida2022sq}. Group residual quantization was introduced in \cite{yang2023hifi}, reducing the number of quantizers required for high-quality reconstruction. In \cite{liu2024semanticodec}, the encoding was separated into semantic and acoustic components, enabling operation at very low bitrates and facilitating language model integration. The Mimi codec \cite{defossez2024moshi} augmented RVQ with a parallel quantizer to distill semantic information, improving phonetic discriminability. 

Beyond achieving high compression rates, neural audio codec representations have proven valuable for downstream tasks. The utility of discrete latents was first demonstrated in computer vision, where convolutional models trained on VQ-VAE representations enabled high-quality image generation \cite{razavi2019generating}. Extending this principle to audio has motivated task-specific codec designs that enforce disentangled and semantically meaningful representations.
Prior work has explored explicit latent disentanglement in neural codecs, separating attributes such as pitch, speaker identity, timbre, vocal effort and prosody for tasks including singing voice conversion, zero-shot speech generation, and speech resynthesis \cite{takahashi2021hierarchical,wang2023generalizable,polyak2021speech, jacquelin2025lombardtokenizer}. Other works instead impose disentanglement through auxiliary objectives, e.g., for speech separation \cite{omran2023disentangling} or via pretext tasks and gradient-reversal tricks for zero-shot speech synthesis \cite{ju2024naturalspeech}.
More general approaches aim to build codecs that support multiple data modalities and subtasks, by separating speech, music, and environmental sounds \cite{bie2025learning,jiang2025unicodec}, or disentangling frequency bands \cite{luo2024gull,ginies2025soft}.

\subsection{Language models for audio application}

Following the success of transformer-based language models \cite{vaswani2017attention,devlin2019bert}, a growing line of work treats discrete audio representations as tokens for language modeling. \cite{baevski2020wav2vec} demonstrated this approach by combining a transformer-based language model with masked encoder latents and contrastive learning, yielding robust discrete audio representations. Building on this idea, masking strategies were integrated in \cite{huang2022masked} with transformer blocks and neural audio codecs to construct latent representations well-suited for classification tasks. Beyond representation learning, language models have also been shown to be effective for generative audio modeling. For instance, in \cite{wang2023neural}, codec-derived discrete units were leveraged with language models for zero-shot text-to-speech, an idea later extended to speech translation \cite{zhang2023speak}. Similarly, in \cite{copet2023simple}, next-token prediction was applied to music generation, further underscoring the generality of this paradigm. A related strategy has recently been explored in speech restoration, where generative language models are trained to predict clean codec tokens from their degraded versions \cite{li2024masksrmaskedlanguagemodel, yang2024genhancer}.

\subsection{Bandwidth extension}

Bandwidth extension, which consists in inferring high-frequency content from low-pass signals, has been studied extensively. Classical approaches focused on spectral manipulation, such as duplicating or rescaling low-frequency spectra into higher bands \cite{Dietz2002SpectralBR,nagel2009harmonic}. Neural methods substantially reshaped the problem, with early applications of U-Nets for reconstructing truncated signals \cite{kuleshov2017audio}. Diffusion-based approaches further advanced performance, including NU-Wave \cite{lee2021nu,han2022nu}, AudioSR \cite{liu2024audiosr} and UniverSR \cite{choi2026universr}, which reconstruct high-frequency details from waveform or mel-spectrogram inputs. A2SB \cite{kong2025a2sb} similarly employs a diffusion-based process, specifically Schrödinger bridges, to address both bandwidth extension and audio inpainting. 
Concurrently, diffusion processes were coupled with neural audio codecs for token enrichment in speech enhancement \cite{fang2025vector}. Some hybrid methods \cite{grumiaux2023efficient} also combine differentiable digital signal processing with neural networks. In \cite{li2025apollo}, the architecture was based on the codec of \cite{luo2024gull}, replacing a processing step between the encoder and the decoder by a transformer model to predict missing information.

\section{Our approach}

\begin{figure*}
    \centering
    \includegraphics[trim={1.0cm 0 1cm 0},clip,width=0.85\linewidth]{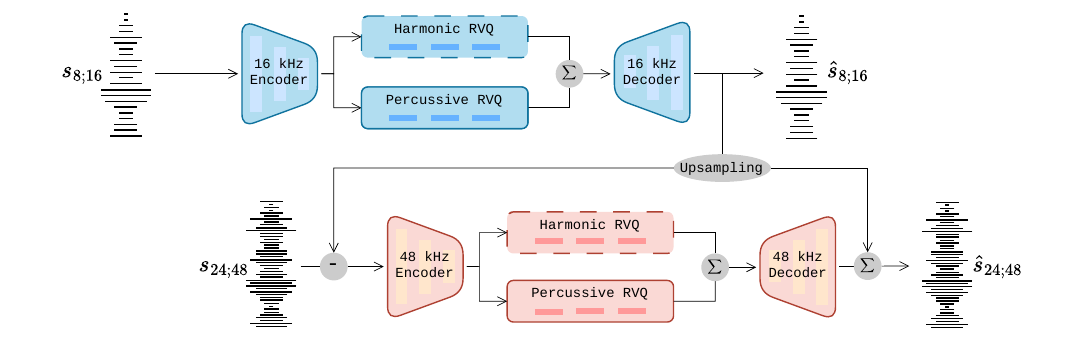}
    \caption{\textbf{HP-codec}, our structure-informed disentangled codec. It is divided in two branches operating at different sampling rates: a 16~kHz branch and a 48~kHz branch. Each branch contains parallel RVQs which are composed of a harmonic section and a percussive section.}
    \vspace{-0.5cm}
\label{fig:codec}
\end{figure*}

\noindent This work is motivated by the strong performance recently demonstrated by transformer-based language models in audio generation tasks, as exemplified by \cite{wang2023neural} and \cite{copet2023simple}. To the best of our knowledge, such approaches have not previously been applied to the bandwidth extension problem, which in itself constitutes a novel contribution of this work.
Beyond this novelty, we argue that a discrete modeling approach is particularly well-suited to bandwidth extension: by reducing the input to a finite set of discrete values, discretization narrows the space of possible outcomes at prediction time, encouraging the model to favor the most probable and parsimonious predictions and thereby mitigating perceptual artifacts. Neural audio codecs provide compact discrete representations well suited to this purpose, enabling the use of NLP-style sequence modeling techniques.
Motivated by these considerations, our work focuses on identifying a discretization scheme capable of faithfully encoding low-frequency information while remaining compatible with the generative structure required for bandwidth extension. \\
We denote by $s$ a time-domain signal, and by $s_{b;SR}$ its version band-limited to $b$ kHz and sampled at the sampling frequency $SR$. With this notation, $s_{8;16}$ corresponds to the signal $s$, band-limited to 8 kHz and sampled at 16 kHz.
Due to the architectural specificities of our approach, we restrict this study to a single bandwidth extension configuration: extending a signal containing information up to 8~kHz to one containing information up to 24~kHz. The goal is thus to reconstruct $s_{24;48}$ from the low-frequency components available in $s_{8;16}$.\\
To this end, we leverage the generative modeling capabilities of transformer-based language models by operating in a discrete token space, which offers a cleaner and more tractable domain in which the transformer can focus on predicting the missing high-frequency structure.\\
We thus propose a two-stage neural architecture that combines a disentangled neural audio codec (HP-codec) with an audio language model to form our bandwidth extension model (HP-codecX). HP-codec is first trained to produce a structured latent representation of the input signal, after which the audio language model is fitted on this latent space to capture and predict the missing high-frequency information. The overall framework is illustrated in Fig.~\ref{fig:codec} and Fig.~\ref{fig:BE}.

\subsection{HP-codec: the Disentangled Codec}

\subsubsection{Frequency disentanglement}
Our codec builds upon the architecture of \cite{ginies2025soft}, itself derived from a mid-to-low-bitrate variant of the DAC codec \cite{kumar2023high}, by introducing a branched design. Specifically, the DAC structure is replicated into two branches: one dedicated to encoding and reconstructing low-frequency components, and the other to high-frequency components. This design enforces a disentanglement of frequency bands in the learned discrete representations, while maintaining a dependency between them. The dependency between frequency bands is enforced by computing the residual between the output of the low-frequency branch and the input to the high-frequency branch, as illustrated in Fig.~\ref{fig:codec}.

In our particular implementation, the first branch operates at a 16~kHz sampling rate and models spectral components up to 8~kHz, while the second branch operates at 48~kHz and captures the remaining content up to 24~kHz. Denoting by $\hat{s}_{8;16}$ the reconstruction produced by the first branch and by $\hat{s}_{8;48}$ its upsampled version at 48~kHz, the input to the second branch is defined as the residual signal $s_{24;48} - \hat{s}_{8;48}$. Since the subsequent bandwidth extension prediction relies on mapping low-frequency branch tokens to high-frequency branch tokens, we select the compression ratios of both branches such that they operate at the same bitrate; this ensures that, for a given audio sample, both branches produce the same number of tokens, with each token corresponding to the same temporal context across branches, thereby simplifying the prediction process. In our setting, each RVQ comprises three consecutive codebooks.

\subsubsection{Structure-informed sections}

To strengthen the spectral structure shared across the two branches of HP-codec, we further decompose the RVQs into two parallel modules: a harmonic RVQ and a percussive RVQ, as shown in Fig.~\ref{fig:codec}. Each module is specialized for encoding harmonic or percussive components of the signal, respectively. 

This choice was motivated by several considerations. First, unlike alternative decomposition strategies that rely on data-specific assumptions (source separation requires prior knowledge of the constituent sources, while speech-specific decompositions are by definition limited to speech), harmonic-percussive (or Harmonic + Noise) decompositions are general-purpose models applicable to virtually any audio signal \cite{serra1990spectral, McAulay92, richard1996analysis, fitzgerald2010harmonic, driedger2014extending}.
Second, and most importantly, this decomposition is particularly well suited to bandwidth extension: harmonic and percussive components, which together account for a substantial portion of the energy in most audio signals, exhibit distinct distributions of energy across time and frequency, with harmonic content appearing as parallel horizontal stripes along a fundamental frequency and its harmonics, and percussive content as impulse-like, Dirac-shaped structures. Once disentangled, these distinct spectral behaviors provide a strong foundation for synthesizing high-frequency content from low-frequency information.
Third, this decomposition also improves the interpretability of our model, providing a clear characterization of the role played by each stage of the process.\\
After quantization, the discrete representations produced by the two sections of each branch are summed and subsequently passed to the decoder, which synthesizes the corresponding time-domain reconstruction.

\subsubsection{Training procedure}
\label{sec:training_procedure}

HP-Codec is trained end-to-end using a composite objective that combines five complementary loss terms: a multi-scale mel-spectrogram reconstruction loss, codebook and commitment losses for quantizer regularization, and adversarial and feature-matching losses computed from multi-period and multi-band multi-scale STFT discriminators, following the discriminator design of \cite{kumar2023high}.

\textbf{Multi-Scale Mel-Spectrogram Loss:}
To encourage spectral fidelity across multiple time-frequency resolutions, we minimize the $L_1$ distance between the log-mel spectrograms of a target signal $s$ and its reconstruction $\hat{s}$:

{\small
\begin{equation*}
\mathcal{L}_{\text{mel}}(s, \hat{s}) = \sum_{i=1}^{K} \Big\| \log_{10}\big(\mathrm{Mel}_i(s)\big) - \log_{10}\big(\mathrm{Mel}_i(\hat{s})\big) \Big\|_{1}
\label{eq:mel_loss}
\end{equation*}
}

where $\mathrm{Mel}_i(\cdot)$ denotes the mel-spectrogram operator at scale $i$, computed over $K = 7$ scales using $\{5, 10, 20, 40, 80, 160, 320\}$ mel bins with corresponding window lengths $\{32, 64, 128, 256, 512, 1024, 2048\}$ (in samples). At each scale, the hop length is set to one-quarter of the window length, and a Hann window is used for the underlying STFT.

\textbf{Codebook and Commitment Losses:}
Let $z_e$ denote the continuous latent representation at the input of the residual vector quantizer (RVQ), and $z_q$ its quantized counterpart. Following standard VQ-VAE training \cite{van2017neural}, we use a commitment loss to encourage the encoder output to stay close to its assigned codebook entry, and a codebook loss to update the codebook toward the encoder output:

{\small
\begin{equation*}
\mathcal{L}_{\text{commit}} = \big\| z_e - \mathrm{sg}[z_q] \big\|_2^2, \qquad
\mathcal{L}_{\text{codebook}} = \big\| z_q - \mathrm{sg}[z_e] \big\|_2^2
\label{eq:vq_losses}
\end{equation*}
}

where $\mathrm{sg}[\cdot]$ denotes the stop-gradient operator, which allows gradients to bypass the non-differentiable quantization step via the straight-through estimator.

\textbf{Adversarial and Feature-Matching Losses:}
Let $s$ denote a real audio signal and $\hat{s}$ its reconstruction. The discriminator, generator, and feature-matching losses are defined as:

{\small
\begin{align*}
\mathcal{L}_{D} &= \sum_{k=1}^{K} \Big[ \mathbb{E}[D_k(\hat{s})^2] + \mathbb{E}[(1-D_k(s))^2] \Big] \\
\mathcal{L}_{G} &= \sum_{k=1}^{K} \mathbb{E}[(1-D_k(\hat{s}))^2] \\
\mathcal{L}_{\text{FM}} &= \sum_{k=1}^{K}\sum_{j=1}^{L_k-1} \| D_k^{(j)}(\hat{s}) - \mathrm{sg}[D_k^{(j)}(s)] \|_1
\end{align*}
}

where $D_k(\cdot)$ denotes the final output of the $k$-th discriminator, $D_k^{(j)}(\cdot)$ its $j$-th intermediate feature map, and $L_k$ the total number of layers (intermediate feature maps plus the final output layer) of discriminator $k$. We employ two families of discriminators, identical in design to \cite{kumar2023high}: (i) multi-period discriminators operating on the raw waveform with periods $\{2, 3, 5, 7, 11\}$, and (ii) multi-band multi-scale STFT discriminators, using window lengths $\{2048, 1024, 512\}$ with a hop length of one-quarter of the window length, each applied over frequency sub-bands defined by the bandlimits $\{0.0, 0.1, 0.25, 0.5, 0.75, 1.0\}$ (as a fraction of the Nyquist frequency).

\textbf{Training principle:}
Training follows the cascade strategy of \cite{ginies2025soft}. We first train the low-frequency branch, then freeze its parameters, and train the high-frequency branch. Finally, we jointly finetune the entire codec. Loss scaling across training phases is also left untouched: DAC’s scheme \cite{kumar2023high} is applied in the first two phases; during finetuning, we aggregate codebook-related losses by summation across branches, while averaging the remaining losses. All phases use an exponential learning-rate scheduler with decay factor $\gamma = 0.999996$, with a base rate of $10^{-4}$ in regular phases and $5\times 10^{-5}$ in finetuning. 

At each training step, we select one of three iteration types $\{\text{harmonic}, \text{percussive}, \text{full}\}$, with equal probability, thereby determining the training objective for that step. In a full iteration, all RVQ sections are updated using batches of non-decomposed signals. In a harmonic iteration, only the harmonic section of each RVQ is trained, with inputs derived from the harmonic component of the harmonic–percussive decomposition~\cite{driedger2014extending}. Conversely, in a percussive iteration, only the percussive section of each RVQ is trained, using inputs derived from the percussive component of the same decomposition. For each branch, the overall codebook and commitment losses are computed as the sum of the corresponding losses across its RVQ sections.

\begin{figure*}
    \centering
    \includegraphics[trim={0.5cm 0 1cm 0},clip,width=0.9\linewidth]{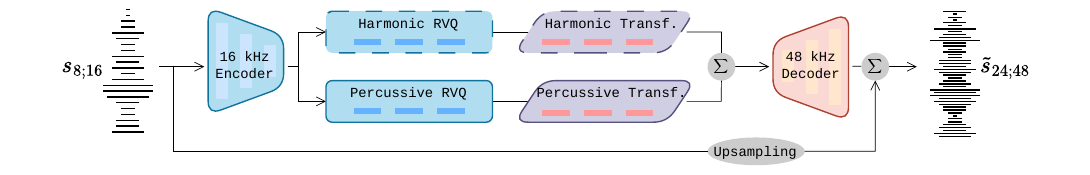}
    \caption{\textbf{HP-codecX}, our bandwidth extension model. It connects the 16~kHz representation, extracted from the input, to the 48~kHz decoder, through an audio language model organized into two sub-models: a harmonic estimator and a percussive estimator.}
    \vspace{-0.5cm}
    \label{fig:BE}
\end{figure*}

\subsection{HP-codecX: Audio language model for prediction}

\subsubsection{Language model}

In the bandwidth extension setting, we only observe $s_{8;16}$, from which we extract the Harmonic and Percussive token sequences from the 16~kHz branch: 
$\{H_n^{16;1}, H_n^{16;2}, H_n^{16;3}\}$ for the first, second and third Harmonic codebooks, and 
$\{P_n^{16;1}, P_n^{16;2}, P_n^{16;3}\}$ for the Percussive codebooks
with $n \in \{1, \dots, N\}$ indexing the tokens within each sequence.

Following \cite{wang2023neural}, we adapt a transformer decoder to operate on the tokens of each RVQ section (Fig.~\ref{fig:BE}). For instance, the harmonic transformer takes as input $\{(H_n^{16;1}),(H_n^{16;2}), (H_n^{16;3})\}$ and predicts $(\Tilde{H}_n^{48;1})$ an estimate of the tokens of the high frequency branch first codebook.
In a second stage, the model uses $\{(H_n^{16;1}),(H_n^{16;2}),(H_n^{16;3}),(\Tilde{H}_n^{48;1})\}$ to predict $(\Tilde{H}_n^{48;2})$. This procedure is applied analogously to the percussive section.

The prediction task is decomposed into two subtasks, yielding the estimated token sequences $\{(\Tilde{H}_n^{48;1}), (\Tilde{H}_n^{48;2}), (\Tilde{H}_n^{48;3})\}$, and $\{(\Tilde{P}_n^{48;1}), (\Tilde{P}_n^{48;2}), (\Tilde{P}_n^{48;3})\}$. These are summed and passed through the 48 kHz decoder to reconstruct the high-frequency components, which are then added to $s_{8;48}$ to produce $\Tilde{s}_{24;48}$, an estimate of the full-band signal $s_{24;48}$.

At inference time, nucleus sampling \cite{holtzman2019curious} is applied with a cumulative probability threshold of 0.95, retaining the smallest set of most probable tokens whose cumulative probability exceeds this value, in order to introduce controlled variability in the generated outputs.

\subsubsection{Training procedure}

All transformer modules of HP-codecX, corresponding to the RVQ sections, are trained using a standard cross-entropy objective and optimized jointly, with the total loss defined as the sum of the cross-entropy terms from each prediction. At each training iteration, the three stages of the prediction process are optimized jointly, with each stage fed the ground-truth tokens relevant to its task (teacher forcing). To convey step-specific information to the transformer, the token embeddings at each step are summed with a learned step embedding prior to being fed to the network. The training uses a cosine annealing learning rate schedule with an initial rate of $10^{-4}$.

\section{Experimental Setup}

\subsection{Baselines}

To benchmark the performance of HP-codecX, we compare it against several state-of-the-art bandwidth extension systems. As our approach to this task remains fairly novel, identifying suitable comparison models proved challenging: existing systems are often either not fully comparable to ours in terms of use case (e.g.,~\cite{grumiaux2023efficient}) or address a broader scope than ours (e.g.,~\cite{kong2025a2sb}). Nonetheless, we selected models capable of performing the exact task considered in this paper. The first is Apollo~\cite{li2025apollo}, which adapts the GULL architecture~\cite{luo2024gull} for this task; it operates at 44.1~kHz and is trained on general-purpose music datasets degraded through low-bitrate MP3 encoding. The remaining baselines (AudioSR~\cite{liu2024audiosr}, UniverSR~\cite{choi2026universr}, and A2SB~\cite{kong2025a2sb}) all perform bandwidth extension via a diffusion process applied to the signal's spectrogram (UniverSR and A2SB are themselves already compared against AudioSR in \cite{liu2024audiosr},~\cite{choi2026universr}).
AudioSR operates at 48~kHz and is trained on music, speech, and environmental sound signals degraded through various low-pass filtering configurations. UniverSR similarly targets 48~kHz reconstruction and is trained on downsampled music, speech, and environmental audio. Finally, A2SB jointly addresses inpainting and bandwidth extension up to 44.1~kHz, trained on downsampled musical audio. Despite differences in the underlying approaches of these models relative to ours, we consider their training conditions sufficiently comparable to support a fair comparison across all baselines.

Regarding the evaluation of HP-codec, our objective was not to design a state-of-the-art codec, but rather to adapt the architecture of DAC \cite{kumar2023high} to the bandwidth extension task while preserving its original strengths as much as possible. Accordingly, we restricted our comparison to retrained 16 kHz and 48 kHz versions of DAC. The 16 kHz variant employs compression ratios of {2,4,5,8}, as suggested by the DAC framework, and encodes audio using 12 codebooks encoded on 10 bits, yielding a bitrate of 6 kbit/s. The 48 kHz DAC model employs compression ratios of {2,4,8,8}, also suggested by the DAC framework, using 12 codebooks encoded on 10 bits as well, for a bitrate of 11.28 kbit/s (the closest achievable match to the 12 kbit/s bitrate of our model). We further compared our model to that proposed in \cite{ginies2025soft}, preserving the compression ratios of HP-codec and the total number of codebooks per branch, in order to verify that our reconstruction performance was not substantially degraded relative to this baseline. In addition, we rely on the comparisons already established in the original DAC paper against other state-of-the-art codecs.

\subsection{Datasets} \label{sec-datasets}
The training of our model has been performed on the training part of the MUSDB18 dataset \cite{stoter20182018} and on the JAMENDO dataset \cite{bogdanov2019mtg}. The testing of HP-codec was performed on the testing part of MUSDB18 dataset \cite{stoter20182018}. The testing of our bandwidth extension model's (HP-codecX) prediction was done on the testing part of the MUSDB18 dataset \cite{stoter20182018}, as well as on datasets that were not observed during training: the ENST-Drums dataset \cite{gillet2006enst}, the OrchideaSOL dataset \cite{cella2020orchideasol}, the Medley-solos-DB dataset \cite{lostanlen2018medley}, and the Monophonic and Polyphonic synthetic datasets which were built according to the implementation designed in \cite{grumiaux2023efficient}. We also tested our model using the VCTK dataset \cite{veaux2013voice} and the ESC-50 dataset \cite{piczak2015esc}.

The JAMENDO and MUSDB18 datasets are music corpora comprising more than 55,000 music signals in their combined training set. Our training set gathers approximately 3,800 hours of music samples, while our test set consists of 1,000 samples randomly extracted from the 50 music tracks of the 3.5-hour-long MUSDB18 test set.

The harmonic–percussive decomposition used during training follows the procedure described in~\cite{driedger2014extending}, which applies horizontal and vertical median filtering to the magnitude spectrogram to obtain estimates of the harmonic and percussive components, respectively. Other harmonic-percussive decomposition methods exist, such as~\cite{duxbury2001separation} and~\cite{ono2008separation}; however, as noted in~\cite{driedger2014extending}, the median filtering approach achieves state-of-the-art separation performance while remaining fast and straightforward to implement.
To obtain strongly separated audio samples, we set the power of the Wiener filters used for masking to 2.0, and the margin for discrimination to 3.0, yielding a highly discriminative filter that ensures the harmonic decomposition predominantly contains harmonic content and the percussive decomposition predominantly contains percussive content. Under this parameterization, the harmonic and percussive components produced by the HPSS procedure do not sum to the original signal, as a residual component is discarded during decomposition. To prevent this loss of information, the full (unseparated) audio samples are retained and included in the training mixture, thereby preserving the content discarded by the HPSS process.

We also constituted testing sets, each composed of 1000 samples extracted from OrchideaSOL and Medley-solos-DB datasets (which gather single instruments recordings), from ENST-Drums dataset (which gathers drums recordings) and from the Monophonic and Polyphonic synthetic datasets (which are composed respectively of purely harmonic sources and superposition of several harmonic sources). The same principle was applied to the speech excerpts of VCTK and the environmental sounds samples of ESC-50. These testing sets were used for out-of-domain testing. \newline
All samples are originally recorded at 44.1~kHz, upsampled to 48~kHz, and thus contain information up to 22.05~kHz, with the exception of the VCTK dataset, whose samples contain information up to 24~kHz.

\subsection{Objective metrics}
\label{sec:obj_metrics}

We assess objective reconstruction quality along three complementary axes: spectral fidelity, waveform fidelity, and perceptual quality.

\textbf{Spectral Metrics:}
Following \cite{kumar2023high}, we adopt multi-resolution mel-spectrogram and STFT metrics to capture spectral discrepancies at multiple time-frequency resolutions.

For a pair of signals $x$ and $y$, the multi-resolution mel metric is defined as:

{\small
\begin{equation*}
\begin{split}
\mathcal{L}_{\text{mel}}^{\text{MR}}(x, y) = \sum_{i=1}^{K} \Biggl[ &\Big\| \log_{10}\big(\mathrm{Mel}_i(x)^{2}\big) - \log_{10}\big(\mathrm{Mel}_i(y)^{2}\big) \Big\|_1\\ 
&+ \Big\| \mathrm{Mel}_i(x) - \mathrm{Mel}_i(y) \Big\|_1 \Biggr]
\end{split}
\label{eq:mel_metric}
\end{equation*}
}

where $\mathrm{Mel}_i(\cdot)$ denotes the mel-spectrogram operator at scale $i$, computed over $K=2$ resolutions using window lengths $\{2048, 512\}$ with corresponding mel-bin counts $\{150, 80\}$.

Analogously, the multi-resolution STFT metric is defined as:

{\small
\begin{equation*}
\mathcal{L}_{\text{stft}}^{\text{MR}}(x, y) = \sum_{i=1}^{K} \Biggl[ \Big\| \log_{10}\big(|X_i|^{2}\big) - \log_{10}\big(|Y_i|^{2}\big) \Big\|_1  + \Big\| |X_i| - |Y_i| \Big\|_1 \Biggr]
\label{eq:stft_metric}
\end{equation*}
}

where $X_i$ and $Y_i$ denote the STFT magnitudes of $x$ and $y$ at scale $i$, computed over the same set of window lengths $\{2048, 512\}$.

\textbf{Waveform Metric:}
To assess sample-level fidelity, we additionally report an $\ell_1$ loss directly on the waveform, $\mathcal{L}_{\ell_1}(x, y) = \| x - y \|_1$.

\textbf{Perceptual Quality:}
As a non-intrusive proxy for perceptual quality, we report ViSQOL scores \cite{chinen2020visqol}.

\textbf{Scale-Invariant Signal-to-Distortion Ratio:}
We further report the scale-invariant signal-to-distortion ratio (SI-SDR) \cite{le2019sdr} as a measure of distortion relative to the underlying content. Given a pair of zero-mean signals $\tilde{x}$ and $\tilde{y}$ (obtained after removing their respective temporal means), the target and residual components are computed as:

{\small
\begin{equation*}
e_{\text{target}} = \frac{\langle \tilde{y}, \tilde{x} \rangle}{\|\tilde{x}\|_2^2} \, \tilde{x}, \qquad e_{\text{res}} = \tilde{y} - e_{\text{target}}
\label{eq:sisdr_components}
\end{equation*}
}

and the SI-SDR is given by:

{\small
\begin{equation*}
\mathcal{L}_{\text{SI-SDR}}(x, y) = -10 \log_{10} \left( \frac{\| e_{\text{target}} \|_2^2}{\| e_{\text{res}} \|_2^2} \right)
\label{eq:sisdr}
\end{equation*}
}

While widely used in audio coding, SI-SDR is less suited to evaluating synthesis-based systems: its sensitivity to sample-level misalignment can penalize outputs that remain perceptually acceptable, as noted in \cite{defossez2024moshi} and \cite{parker2024scaling}.

\textbf{Pivotal bootstrap confidence interval:}
Alongside the objective metrics described above, we report, where the presentation format allows, the associated 95\% bootstrap confidence intervals, computed using the pivotal bootstrap method \cite{tibshirani1993introduction}.

\subsection{Listening test} \label{sec-listening_test}

The perceptual quality of the proposed bandwidth extension method was further assessed through a MUSHRA listening test \cite{schoeffler2018webmushra}. To keep the test short enough for listeners to complete comfortably, we restricted the comparison to two competing systems: Apollo (APO) \cite{li2025apollo} and A2SB (A2S) \cite{kong2025a2sb}. Apollo was selected because it is methodologically distinct from the other diffusion-based baselines considered in this work, while A2SB was selected on the basis of its strong objective performance in bandwidth extension quality, as well as its prior comparison against the AudioSR model \cite{liu2024audiosr}.
Thirteen non-expert listeners took part in the study, conducted under standard office conditions using headphones, with free replay of excerpts permitted throughout. Each listener completed $12$ trials, each comprising $5$ stimuli rated on a continuous $0$--$100$ scale relative to a reference signal. For each input $s$, a trial consisted of the low-pass anchor $s_{8;16}$, the hidden reference $s_{24;48}$, and the three system outputs $\tilde{s}_{24;48}^{\text{APO}}$, $\tilde{s}_{24;48}^{\text{A2S}}$, and $\tilde{s}_{24;48}^{\text{HPX}}$. In accordance with standard MUSHRA practice, the reference signal was also displayed openly and made available for comparison throughout each trial. The $12$ excerpts were randomly drawn from the MUSDB18 test set, with six selected from segments exhibiting high energy in the high-frequency band and six from the remaining segments, so as to probe both favorable and challenging operating conditions.
Listeners were instructed to base their ratings specifically on the fidelity of the high-frequency content relative to the reference, and to assign a score of $100$ to at least one stimulus per trial, since the hidden reference was always present among the five items. Participants who did not assign a sufficiently high score to the hidden reference were excluded from the study, following standard MUSHRA post-screening practice. For each system, we report the median rating together with the first and third quartiles of the collected scores as the perceptual evaluation metric.

\subsection{Technical specifications}

HP-codec adopts the DAC architecture \cite{kumar2023high}, adapted following the framework proposed in \cite{ginies2025soft}, with a branched design operating at 6~kbit/s (16~kHz branch) and 12~kbit/s (48~kHz branch), each using three codebooks per Harmonic/Percussive RVQ section. HP-codecX's transformer modules follow the design of \cite{wang2023neural}, each using a 6-layer decoder with 8 attention heads and hidden dimension 4096. Both models were trained on a single NVIDIA L40S GPU (48~GB), for approximately 20 hours (HP-codec) and 27 hours (HP-codecX). Full architectural and training details (encoder rates, compression ratios, segment lengths, batch sizes, iteration counts) are available on our \href{https://harmonic-percussive-bandwidth-extension.github.io/}{\underline{companion website}}.

\section{Testing HP-codec}
\subsection{Validating architectural choices}

To verify that introducing structure-informed sections does not degrade the reconstruction quality of HP-codec, we evaluate the model on the MUSDB18 \cite{stoter20182018} test set. For comparison, we adapt the disentangled codec of \cite{ginies2025soft} (soft-dis-DAC) to operate at 48 kHz under identical compression rates, and we retrain a 16~kHz and a 48~kHz DAC models \cite{kumar2023high} at the same compression ratios. Since DAC is a widely used baseline with extensive comparisons in the literature, including it provides a clearer sense of how HP-codec aligns with existing methods. The results are summarized in Table \ref{tab-codec}.

\begin{table}[h]
\caption{Reconstruction metrics (and 95\% CI) for HP-codec.\\ \normalfont\itshape{} Soft-dis-DAC is a modified version of \cite{ginies2025soft} in which the structure-informed sections (Harmonic and Percussive RVQs) are removed. DAC-16kHz and DAC-48kHz denotes DAC models \cite{kumar2023high} retrained on our dataset.}
\begin{center}

\scalebox{0.8}{
\begin{tabular}{|c|c|c|c|c|}
\multicolumn{2}{c}{}&\multicolumn{3}{c}{\textbf{$SR = 16~kHz$}} \\
\noalign{\vskip 6pt}
\cline{3-5}
\multicolumn{2}{c|}{} & \thead{Mel $\downarrow$} & \thead{STFT $\downarrow$} & \thead{Waveform $\downarrow$\\($10^{-2}$)} \\
\cline{1-1} \cline{3-5}
HP-codec&& 0.79 {\tiny [0.77,0.78]}& 2.26 {\tiny [2.25,2.28]}& 4.8 {\tiny [4.8,4.9]} \\
soft-dis-DAC \cite{ginies2025soft}&& \cellcolor{gray!30}0.70 {\tiny [0.69,0.70]}& \cellcolor{gray!30}2.11 {\tiny [2.10,2.13]}& \cellcolor{gray!30}4.1 {\tiny [4.1,4.2]} \\
DAC \cite{kumar2023high}&& 0.82 {\tiny [0.81,0.82]}& 2.32 {\tiny [2.30,2.34]}& 5.1 {\tiny [5.0,5.2]} \\
\cline{1-1} \cline{3-5}
\end{tabular}
}

\vspace{6pt}

\scalebox{0.8}{
\begin{tabular}{|c|c|c|c|}
\cline{3-4}
\multicolumn{2}{c|}{} & \thead{SI-SDR $\uparrow$} & \thead{ViSQOL $\uparrow$} \\
\cline{1-1} \cline{3-4}
HP-codec&& 7.23 {\tiny [7.0,7.4]}& 4.34 {\tiny [4.33,4.34]} \\
soft-dis-DAC \cite{ginies2025soft}&& \cellcolor{gray!30}8.75 {\tiny [8.57,8.93]}& \cellcolor{gray!30}4.43 {\tiny [4.42,4.43]} \\
DAC \cite{kumar2023high}&& 6.80 {\tiny [6.60,6.99]}& 4.37 {\tiny [4.37,4.38]} \\
\cline{1-1} \cline{3-4}
\end{tabular}
}

\vspace{12pt}

\scalebox{0.8}{
\begin{tabular}{|c|c|c|c|c|}
\multicolumn{2}{c}{}&\multicolumn{3}{c}{\textbf{$SR = 48~kHz$}} \\
\noalign{\vskip 6pt}
\cline{3-5}
\multicolumn{2}{c|}{} & \thead{Mel $\downarrow$} & \thead{STFT $\downarrow$} & \thead{Waveform $\downarrow$\\($10^{-2}$)} \\
\cline{1-1} \cline{3-5}
HP-codec&& 0.77 {\tiny [0.77,0.78]}& 2.26 {\tiny [2.25,2.28]}& 5.0 {\tiny [4.9,5.1]} \\
soft-dis-DAC \cite{ginies2025soft}&& \cellcolor{gray!30}0.72 {\tiny [0.72,0.73]}& \cellcolor{gray!30}2.22 {\tiny [2.20,2.23]}& 4.3 {\tiny [4.2,4.4]} \\
DAC \cite{kumar2023high}&& 0.75 {\tiny [0.75,0.76]}& 2.24 {\tiny [2.22,2.26]}& \cellcolor{gray!30}4.1 {\tiny [4.0,4.2]} \\
\cline{1-1} \cline{3-5}
\end{tabular}
}

\vspace{6pt}

\scalebox{0.8}{
\begin{tabular}{|c|c|c|c|}
\cline{3-4}
\multicolumn{2}{c|}{} & \thead{SI-SDR $\uparrow$} & \thead{ViSQOL $\uparrow$} \\
\cline{1-1} \cline{3-4}
HP-codec&& 6.75 {\tiny [6.59,6.92]}& \cellcolor{gray!30}4.36 {\tiny [4.35,4.37]} \\
soft-dis-DAC \cite{ginies2025soft}&& 8.10 {\tiny [8.28,7.92]}& 4.33 {\tiny [4.32,4.34]} \\
DAC \cite{kumar2023high}&& \cellcolor{gray!30}8.79 {\tiny [8.61,8.97]}& 3.92 {\tiny [3.90,3.94]} \\
\cline{1-1} \cline{3-4}
\end{tabular}
}

\label{tab-codec}
\end{center}
\end{table}

These results demonstrate that modifying the RVQ structure to produce a more structure-informed discrete representation in HP-codec yields performance competitive with both the soft disentangled reference model and the retrained DAC baseline.

\subsection{Exploring the Harmonic-Percussive disentanglement}

In this work, we proposed a disentanglement strategy for the representations extracted by a neural audio codec. Our approach modifies the structure of each RVQ in the model and leverages a training procedure inspired by harmonic–percussive decomposition \cite{fitzgerald2010harmonic, driedger2014extending} to enforce disentanglement. Within this framework, each RVQ section is designed to capture a distinct spectral property of the input: the harmonic section encodes harmonic components and the percussive section captures percussive events.

Building on the harmonic–percussive decomposition algorithm, we consider two signal components: harmonic and percussive. Based on this, we designed an experiment in which HP-codec was provided with three types of input: full signals (\textbf{Global}), harmonic components (\textbf{H}) and percussive components (\textbf{P}). Reconstructions were then evaluated under three corresponding decoding settings, where only the relevant RVQ sections (\textbf{Global}, \textbf{H}, or \textbf{P}) were used. This setup yields $3 \times 3 = 9$ input–reconstruction pairs.

\begin{figure}
    \centering
    \includegraphics[trim={0cm 0cm 0cm 0cm},clip,width=\linewidth]{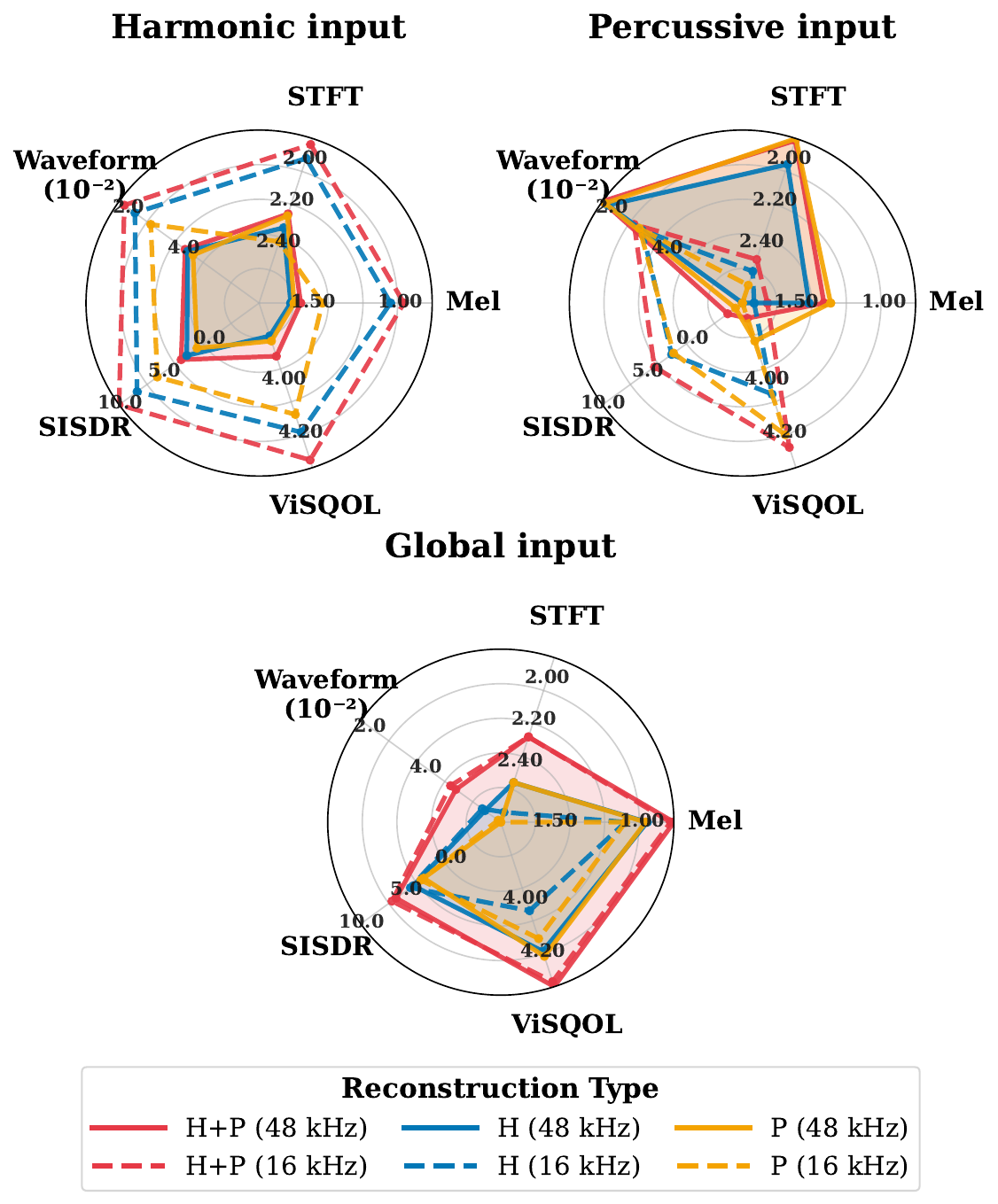}
    \caption{Reconstruction metrics for HP-codec as a function of the spectral composition of the input and the RVQ sections used for reconstruction: harmonic tokens (\textbf{H}), percussive tokens (\textbf{P}), or both (\textbf{H+P}). These plots are to be read as follows: the larger the enclosed area, the better the reconstruction. The Global configuration (\textbf{H+P}) consistently performs best, reflecting its larger representational capacity. The Harmonic section achieves the best performance on harmonic inputs at 16~kHz, while the Percussive section performs best on percussive inputs at 48~kHz. This pattern mirrors the natural spectral distribution of audio, with harmonic energy concentrated at low frequencies and percussive energy extending toward higher frequencies.}
    \vspace{-0.5cm}
    \label{fig:A-spider-codec}
\end{figure}

The results of this experiment are reported in Fig.~\ref{fig:A-spider-codec}. Beyond the expected advantage of the Global configuration, an interesting pattern emerges for section-specific reconstructions: percussive inputs reconstructed with the global configuration show limitations in the 16~kHz branch but are refined in the 48~kHz branch, and for this type of input, the percussive section produces the most accurate reconstruction, whereas for harmonic inputs at 16~kHz, the harmonic section performs best. These findings align with the spectral distribution of natural audio: harmonic components dominate at low frequencies, while percussive components retain significant energy at higher frequencies.

To properly assess the disentanglement of Harmonic and Percussive information between the sections, we conducted the following experiment.

Let $\mathcal{HPSS}_{H}(s)$ and $\mathcal{HPSS}_{P}(s)$ denote the Harmonic and Percussive signals resulting from the HPSS decomposition of a signal $s$. For a signal $s_{f;f'}$, let $\hat{s}_{f;f'}$ denote the reconstruction produced by HP-codec, and let $\hat{s}^{H}_{f;f'}$ and $\hat{s}^{P}_{f;f'}$ denote the reconstructions produced by HP-codec using only the Harmonic tokens or only the Percussive tokens, respectively.

Further, let $E[s] = \sum_{n=-\infty}^{+\infty}|s(n)|^2$ denote the energy of a signal $s$. We then define:

{\small
\begin{equation*}
R_H(s) = \frac{E[\mathcal{HPSS}_H(s)]}{E[s]}, \qquad R_P(s) = \frac{E[\mathcal{HPSS}_P(s)]}{E[s]},
\end{equation*}
}

which respectively correspond to the proportion of Harmonic and Percussive energy contained in a signal $s$.

The experiment consisted in measuring, on the test set, the ratios:

{\small
\begin{equation*}
\frac{R_H(\hat{s}^H_{f;f'})}{R_H(\hat{s}_{f;f'})}, \quad \frac{R_P(\hat{s}^H_{f;f'})}{R_P(\hat{s}_{f;f'})}, \quad \frac{R_H(\hat{s}^P_{f;f'})}{R_H(\hat{s}_{f;f'})}, \quad \frac{R_P(\hat{s}^P_{f;f'})}{R_P(\hat{s}_{f;f'})}
\end{equation*}
}

for $(f,f') = (8,16)_{{kHz}}$ and $(f,f') = (24,48)_{{kHz}}$. 
The underlying idea is to measure the extent to which harmonic energy is more concentrated in the harmonic reconstruction, and percussive energy more concentrated in the percussive reconstruction. 

\begin{figure}
    \centering
    \includegraphics[trim={0cm 0cm 0cm 0cm},clip,width=0.85\linewidth]{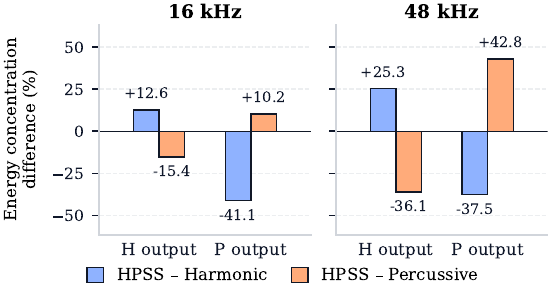}
    \caption{Variation in harmonic and percussive energy concentration between reconstructions using only harmonic (\textbf{H}) or only percussive (\textbf{P}) tokens and the full reconstruction (\textbf{H+P}).}
    \vspace{-0.5cm}
    \label{fig:disentanglement_comp}
\end{figure}

The results, gathered in Fig.~\ref{fig:disentanglement_comp}, clearly show that the Harmonic reconstruction (using only the Harmonic tokens) contains a greater proportion of Harmonic energy than the full reconstruction (using all tokens), for both the 16~kHz and 48~kHz cases. The same observation holds for the Percussive energy in the Percussive reconstruction. These results thus confirm the effective disentanglement of Harmonic and Percussive information within HP-codec.

\subsection{Modulating the Harmonic-Percussive mix}

A direct application of the harmonic-percussive disentanglement learned in the latent representation is the ability to modulate the harmonic-percussive balance of the reconstructed signal. We conduct an experiment in which the outputs of the two RVQs within each frequency branch are linearly weighted prior to decoding: the harmonic scale is varied from 0.0 (no harmonic contribution) to 1.0 (full harmonic contribution) while the percussive scale is held fixed at 1.0, and symmetrically for the percussive scale while the harmonic scale is held fixed. This progressively increases the relative contribution of the harmonic (resp. percussive) component in the final mix. 
Audio samples on our \href{https://harmonic-percussive-bandwidth-extension.github.io/}{\underline{companion website}} illustrate the fine-grained controllability our codec offers over the harmonic-percussive mix. 

\section{Testing HP-codecX}
\subsection{Quality of the bandwidth extension}
\label{sec-language_model}

\begin{figure}
    \centering
    \vspace{-0.1cm}
    \includegraphics[width=1\linewidth]{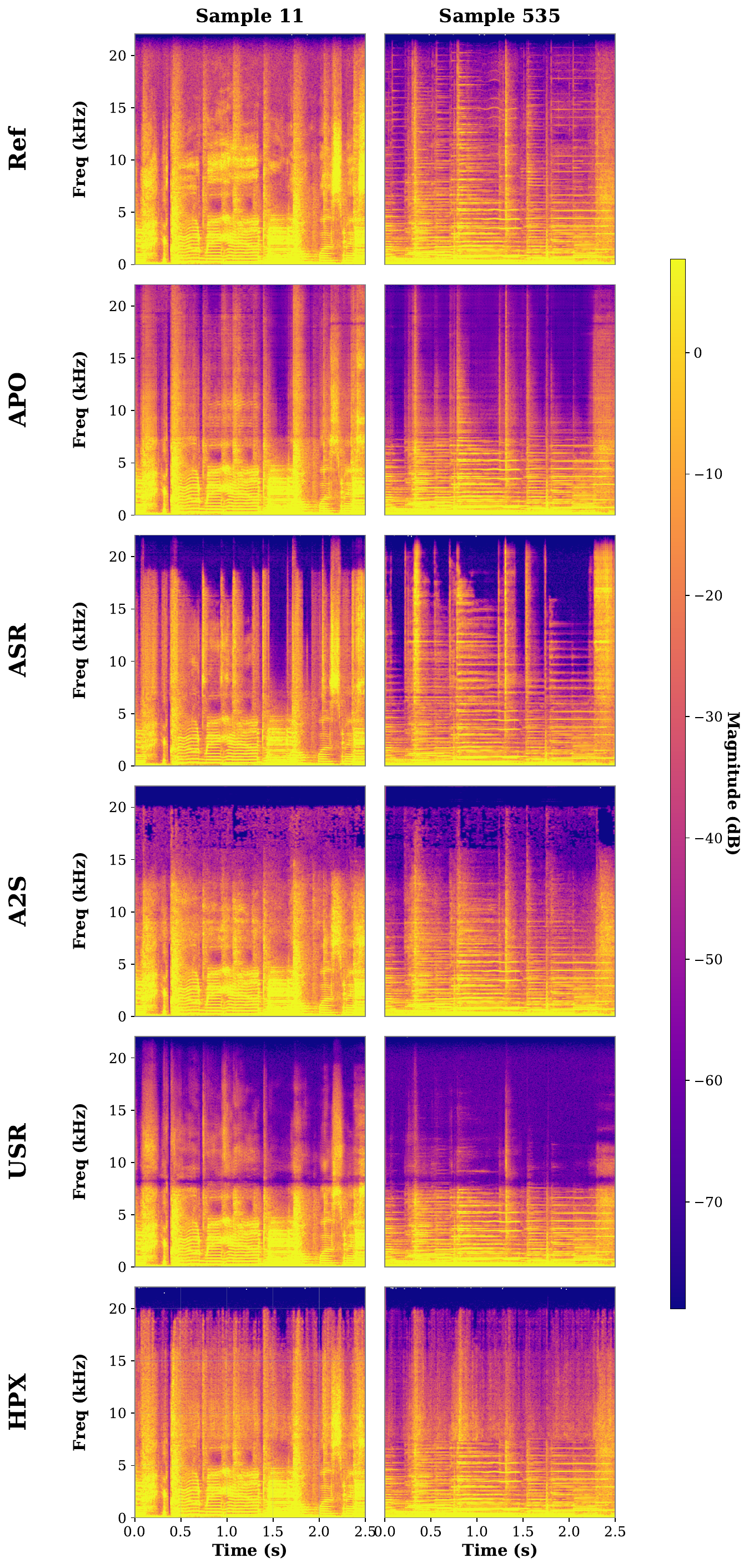}
    \caption{Spectrograms of estimated signal drawn from Apollo (\textbf{APO}), AudioSR (\textbf{ASR}), A2SB (\textbf{A2S}), UniverSR (\textbf{USR}), and HP-codecX (\textbf{HPX}).}
    \vspace{-0.4cm}
    \label{fig:A_recons_spec_1}
\end{figure}

\begin{figure}
    \centering
    \scalebox{0.65}{    \includegraphics[width=0.8\linewidth]{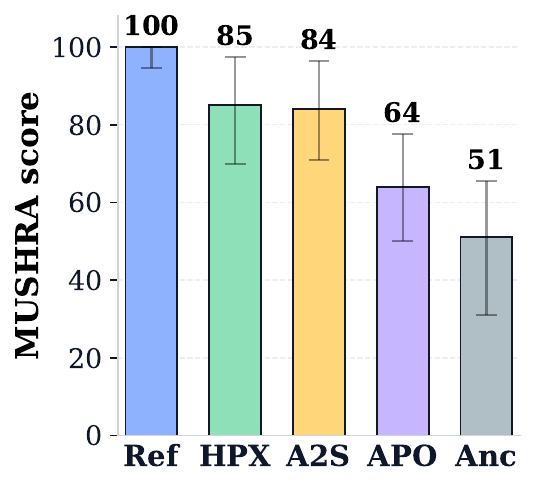}}
    \caption{Results of the perceptual evaluation (with median, first and third quartiles). The MUSHRA test compared Apollo (\textbf{APO}) and A2SB (\textbf{A2S}) models to HP-codecX (\textbf{HPX}). Reference signals (\textbf{Ref}) and anchor signals (\textbf{Anc}) were also evaluated.}
    \vspace{-0.4cm}
    \label{fig:MUSHRA}
\end{figure}

We assessed the quality of the bandwidth extension over the estimated signals using the reconstruction metrics introduced previously, comparing HP-codecX against Apollo (APO), AudioSR (ASR), UniverSR (USR) and A2SB(A2S). Table~\ref{tab-BE-obj} reports results estimations performed on MUSDB18 test set \cite{stoter20182018}. Examples of spectrograms of estimated signals are displayed in Fig.~\ref{fig:A_recons_spec_1}. Fig.~\ref{fig:MUSHRA} gathers the results of the perceptual test introduced in Section \ref{sec-listening_test}. Audio examples are given on our \href{https://harmonic-percussive-bandwidth-extension.github.io/}{\underline{companion website}}.

\begin{table}
\caption{Objective bandwidth extension metrics (and 95\% CI)\\ \normalfont\itshape{} for Apollo (44.1~kHz), AudioSR (48~kHz), A2SB (44.1~kHz), UniverSR (48~kHz) and HP-codecX (48~kHz) on MUSDB18 test set.}
\centering
\scalebox{0.8}{
\begin{tabular}{|l|c|c|c|c|}
\cline{3-5}
\multicolumn{2}{c|}{}& \thead{Mel $\downarrow$} & \thead{STFT $\downarrow$} & \thead{Waveform $\downarrow$\\($10^{-2}$) } \\
\cline{1-1} \cline{3-5}
Apollo \cite{li2025apollo}    && 0.84 {\tiny [0.83,0.85]} & 2.44 {\tiny [2.41,2.47]} & 4.8 {\tiny [4.8,4.9]} \\
AudioSR \cite{liu2024audiosr} && 1.77 {\tiny [1.74,1.80]} & 3.60 {\tiny [3.55,3.64]} & 6.9 {\tiny [6.8,7.0]} \\
A2SB \cite{kong2025a2sb}      && 0.63 {\tiny [0.62,0.64]} & 2.65 {\tiny [2.61,2.69]} & 1.2 {\tiny [1.2,1.3]} \\
UniverSR \cite{choi2026universr} && 1.32 {\tiny [1.29,1.34]} & 3.53 {\tiny [3.48,3.58]} & \cellcolor{gray!30}0.9 {\tiny [0.9,1.0]} \\
HP-CodecX && \cellcolor{gray!30}0.45 {\tiny [0.44,0.46]} & \cellcolor{gray!30}1.89 {\tiny [1.87,1.91]} & 1.2 {\tiny [1.2,1.3]} \\
\cline{1-1} \cline{3-5}
\end{tabular}
}

\vspace{6pt}

\scalebox{0.8}{
\begin{tabular}{|l|c|c|c|}
\cline{3-4}
\multicolumn{2}{c|}{} & \thead{SI-SDR $\uparrow$\\ } & \thead{ViSQOL $\uparrow$} \\
\cline{1-1} \cline{3-4}
Apollo \cite{li2025apollo}    && 3.23 {\tiny [3.00,3.47]} & 3.23 {\tiny [3.20,3.27]} \\
AudioSR \cite{liu2024audiosr} && 13.75 {\tiny [13.44,14.05]} & 3.06 {\tiny [3.03,3.08]} \\
A2SB \cite{kong2025a2sb}      && 19.70 {\tiny [19.36,20.04]} & 3.08 {\tiny [3.05,3.11]} \\
UniverSR \cite{choi2026universr} && \cellcolor{gray!30}22.38 {\tiny [21.94,22.80]} & 2.28 {\tiny [2.25,2.31]} \\
HP-CodecX && 20.23 {\tiny [19.78,20.66]} & \cellcolor{gray!30}3.33 {\tiny [3.31,3.36]} \\
\cline{1-1} \cline{3-4}
\end{tabular}
}
\label{tab-BE-obj}
\end{table}

HP-codecX achieves the lowest Mel and STFT distances among all baselines (Table~\ref{tab-BE-obj}), indicating superior reconstruction of high-frequency spectral content.

\begin{figure*}
    \centering
    \hspace{-1.5cm}
    \scalebox{1}{
    \includegraphics[trim={0cm 0cm 0cm 0cm},clip,width=0.9\linewidth]{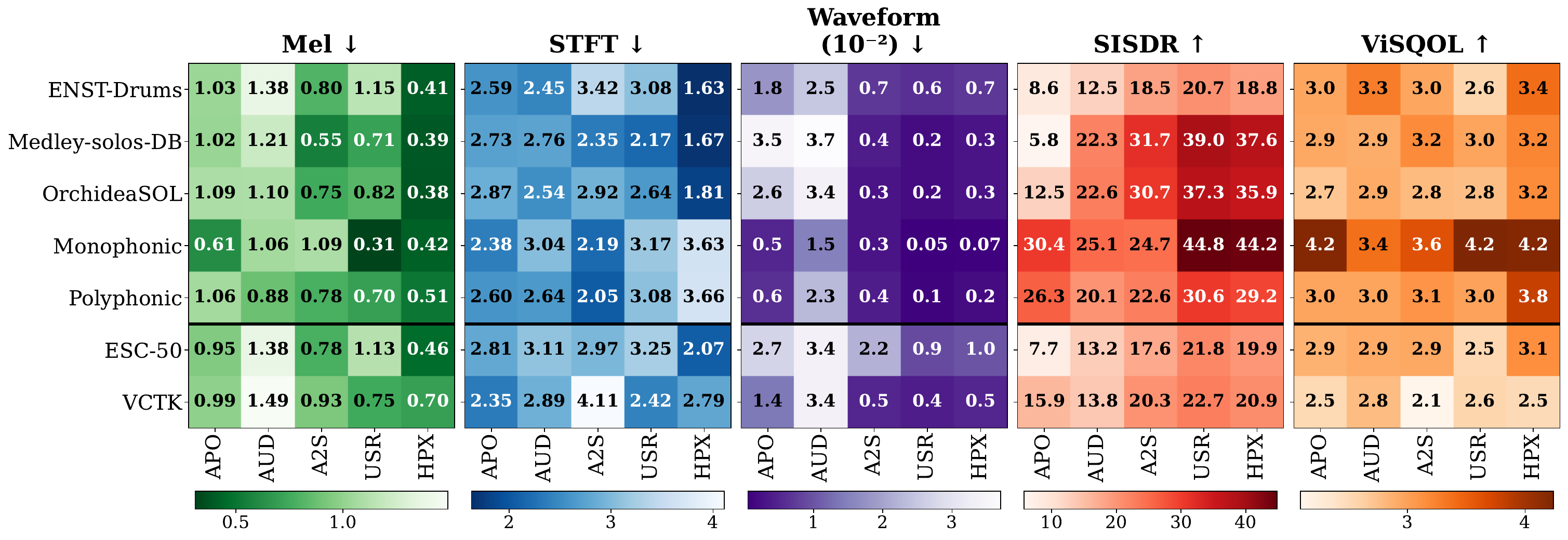}
    }
    \caption{Objective reconstruction metrics calculated on estimated signals. These metrics have been computed on Out-of-Domain test datasets: ENST-Drums, Medley-solos-DB, OrchideaSOL, Monophonic, Polyphonic, ESC-50 and VCTK. The Apollo (\textbf{APO}) and A2SB (\textbf{A2S}) metrics are calculated at 44.1~kHz, while the AudioSR (\textbf{ASR}), UniverSR (\textbf{USR}) and HP-codecX (\textbf{HPX}) metrics have been calculated at 48~kHz.}
    \label{fig:A-Heatmap-LM}
    \vspace{-0.5cm}
\end{figure*}

An evaluation was conducted on the additional testing datasets, using the two baselines and our model, giving the results plotted in Fig.~\ref{fig:A-Heatmap-LM}. This experiment evaluates the bandwidth extension quality of all models on previously unseen data types, using out-of-domain datasets.

The results reveal a clear spectral advantage of HP-codecX on percussive sources (ENST-Drums) and more general music signals (Medley-solos-DB, OrchideaSOL). This advantage diminishes for purely harmonic signals (Monophonic, Polyphonic), although our model remains competitive overall. Consistent with the objective results, informal listening tests confirmed that HP-codecX excels at reconstructing percussive and noise-like components relative to the baselines.

Despite being trained exclusively on music data, our model also achieves strong results on speech (VCTK) and environmental audio (ESC-50). This cross-domain performance supports the robustness of the proposed approach and highlights its capacity to generalize beyond the training distribution.

\subsection{Ablation study}

To validate the architectural and processing choices adopted in this work, we conducted an ablation study of HPcodecX within the bandwidth extension framework. Four design choices central to the proposed architecture were identified and iteratively removed from the model in order to assess their individual contribution to bandwidth extension performance. These design choices are defined as follows:

\begin{itemize}
\item \textbf{Structure-informed training}: the model is trained on a combination of full audio samples and Harmonic/Percussive components obtained through HPSS decomposition.
\item \textbf{Multi-RVQ}: the residual vector quantizers (RVQs) are partitioned into two sections, respectively associated with the Harmonic and Percussive components of the signal.
\item \textbf{Multi-transformer prediction}: two transformers are used instead of one, each dedicated to predicting the tokens of one RVQ subset (Harmonic or Percussive), in accordance with the Multi-RVQ structure.
\item \textbf{Frequency branches}: the modification introduced in \cite{ginies2025soft}, where high- and low-frequency components are separated within the latent representation.
\end{itemize}

Based on the successive removal of these components, five alternative configurations of HPcodecX were derived for bandwidth extension. In all cases, the bitrate was held constant across configurations by adjusting the number of quantizers, ensuring a fair comparison at a fixed compression ratio. The resulting performance is reported in Table~\ref{tab-wosem}. All configurations follow the same experimental protocol, with the exception of EXP5, for which a standard 48~kHz DAC model was employed. In this case, downsampled audio was provided as input, yielding degraded tokens at the encoding stage; a transformer was subsequently trained to reconstruct full-resolution tokens from these degraded representations.

\begin{table}
\caption{Ablation Study: Objective Bandwidth Extension Metrics\\ (and 95\% CI)\\ \normalfont\itshape{} for \textbf{HP-codecX} and five additional configurations obtained by progressively removing architectural and design choices introduced in this work. \textbf{EXP~5} corresponds to a configuration in which a simple DAC model is used to predict full-bandwidth tokens directly from lossy tokens.}
\centering

\scalebox{0.8}{
\begin{tabular}{|l|c|c|c|c|}
\cline{3-5}
 \multicolumn{2}{c|}{}& \thead{HP-CodecX} & \thead{EXP~1} & \thead{EXP~2} \\
\cline{1-1} \cline{3-5}
Spec. informed Training    && \checkmark & $\times$ & $\times$ \\
Multi-transformer     && \checkmark & \checkmark & $\times$ \\
Multi-RVQ             && \checkmark & \checkmark & \checkmark \\
Frequency Branches    && \checkmark & \checkmark & \checkmark \\
\cline{1-1} \cline{3-5}
Mel $\downarrow$      && \cellcolor{gray!30}0.45 {\tiny [0.44,0.46]} & 0.48 {\tiny [0.47,0.48]} & 0.48 {\tiny [0.47,0.48]} \\
STFT $\downarrow$     && \cellcolor{gray!30}1.89 {\tiny [1.87,1.91]} & 1.97 {\tiny [1.92,1.99]} & 1.97 {\tiny [1.95,1.99]} \\
Waveform $\downarrow$ ($10^{-2})$ && \cellcolor{gray!30}1.2 {\tiny [1.2,1.3]} & \cellcolor{gray!30}1.2 {\tiny [1.1,1.2]} & \cellcolor{gray!30}1.2 {\tiny [1.1,1.2]} \\
SI-SDR $\uparrow$     && \cellcolor{gray!30}20.23 {\tiny [19.78,20.66]} & 20.03 {\tiny [19.59,20.44]} & 20.00 {\tiny [19.57,20.41]} \\
ViSQOL $\uparrow$     && \cellcolor{gray!30}3.33 {\tiny [3.31,3.36]} & 3.08 {\tiny [3.05,3.10]} & 3.07 {\tiny [3.04,3.09]} \\
\cline{1-1} \cline{3-5}
\end{tabular}
}

\vspace{6pt}

\scalebox{0.8}{
\begin{tabular}{|l|c|c|c|c|}
\cline{3-5}
 \multicolumn{2}{c|}{}& \thead{EXP~3} & \thead{EXP~4} & \thead{EXP~5} \\
\cline{1-1} \cline{3-5}
Spec. informed Training    && $\times$ & \checkmark & $\times$ \\
Multi-transformer     && $\times$ & $\times$ & $\times$ \\
Multi-RVQ            && $\times$ & $\times$ & $\times$ \\
Frequency Branches    && \checkmark & \checkmark & $\times$ \\
\cline{1-1} \cline{3-5}
Mel $\downarrow$      && 0.84 {\tiny [0.83,0.85]} & 0.58 {\tiny [0.57,0.59]} & 1.06 {\tiny [1.05,1.07]} \\
STFT $\downarrow$     && 2.83 {\tiny [2.80,2.86]} & 2.24 {\tiny [2.21,2.27]} & 2.81 {\tiny [2.79,2.84]} \\
Waveform $\downarrow$ ($10^{-2}$) && 2.4 {\tiny [2.3,2.5]} & 1.5 {\tiny [1.4,1.6]} & 5.7 {\tiny [5.6,5.8]} \\
SI-SDR $\uparrow$     && 14.63 {\tiny [14.17,15.08]} & 18.58 {\tiny [18.14,19.01]} & 6.68 {\tiny [6.51,6.85]} \\
ViSQOL $\uparrow$     && 2.48 {\tiny [2.45,2.51]} & 2.56 {\tiny [2.53,2.59]} & 3.14 {\tiny [3.11,3.17]} \\
\cline{1-1} \cline{3-5}
\end{tabular}
}

\label{tab-wosem}
\end{table}

Removing any of the four components degrades performance monotonically (Table~\ref{tab-wosem}), confirming their individual contributions

\begin{table}
\caption{Objective Bandwidth Extension Metrics (and 95\% CI) \\ \normalfont\itshape{} across various datasets, using only harmonic (\textbf{H}) tokens, only percussive (\textbf{P}) tokens, or the combination of both (\textbf{H+P}).}
\centering

\scalebox{0.8}{
\begin{tabular}{|c|c|cc|c|cc|}
\multicolumn{2}{c}{} & \multicolumn{2}{c}{\textit{General Music}} & \multicolumn{1}{c}{}& \multicolumn{2}{c}{\textit{Harmonic}} \\
\cline{3-4} \cline{6-7}
\multicolumn{2}{c|}{} & Medley-solos-DB & OrchideaSOL & \multicolumn{1}{c|}{}& Monophonic & Polyphonic \\
\cline{3-4} \cline{6-7}
\multicolumn{2}{c}{}&\multicolumn{5}{c}{\textbf{Mel} $\downarrow$} \\
\cline{1-1} \cline{3-4} \cline{6-7}
H   & & 0.42 {\tiny [0.41,0.43]} & 0.40 {\tiny [0.39,0.41]} & & \cellcolor{gray!30}0.41 {\tiny [0.39,0.42]} & \cellcolor{gray!30}0.51 {\tiny [0.50,0.52]} \\
P   & & \cellcolor{gray!30}0.41 {\tiny [0.40,0.42]} & \cellcolor{gray!30}0.39 {\tiny [0.38,0.40]} & & 0.44 {\tiny [0.43,0.46]} & 0.53 {\tiny [0.52,0.53]} \\
\cline{1-1} \cline{3-4} \cline{6-7}
H+P & & 0.39 {\tiny [0.39,0.40]} & 0.38 {\tiny [0.37,0.39]} & & 0.42 {\tiny [0.41,0.44]} & 0.51 {\tiny [0.50,0.52]} \\
\cline{1-1} \cline{3-4} \cline{6-7}
\multicolumn{2}{c}{}&\multicolumn{5}{c}{\textbf{STFT} $\downarrow$} \\
\cline{1-1} \cline{3-4} \cline{6-7}
H   & & \cellcolor{gray!30}1.71 {\tiny [1.69,1.74]} & 1.87 {\tiny [1.83,1.91]} & & \cellcolor{gray!30}3.52 {\tiny [3.47,3.58]} & \cellcolor{gray!30}3.56 {\tiny [3.53,3.59]} \\
P   & & \cellcolor{gray!30}1.71 {\tiny [1.69,1.74]} & \cellcolor{gray!30}1.84 {\tiny [1.80,1.88]} & & 3.68 {\tiny [3.63,3.73]} & 3.73 {\tiny [3.70,3.76]} \\
\cline{1-1} \cline{3-4} \cline{6-7}
H+P & & 1.67 {\tiny [1.65,1.70]} & 1.81 {\tiny [1.77,1.86]} & & 3.63 {\tiny [3.58,3.68]} & 3.66 {\tiny [3.63,3.69]} \\
\cline{1-1} \cline{3-4} \cline{6-7}
\multicolumn{2}{c}{}&\multicolumn{5}{c}{\textbf{Waveform} $\boldsymbol{\downarrow}$ \textbf{(}$\boldsymbol{10^{-2}}$\textbf{)}} \\
\cline{1-1} \cline{3-4} \cline{6-7}
H   & & \cellcolor{gray!30}0.3 {\tiny [0.2,0.3]} & \cellcolor{gray!30}0.3 {\tiny [0.3,0.3]} & & \cellcolor{gray!30}0.08 {\tiny [0.07,0.08]} & \cellcolor{gray!30}0.2 {\tiny [0.2,0.2]} \\
P   & & \cellcolor{gray!30}0.3 {\tiny [0.2,0.3]} & \cellcolor{gray!30}0.3 {\tiny [0.3,0.3]} & & \cellcolor{gray!30}0.07 {\tiny [0.07,0.08]} & \cellcolor{gray!30}0.2 {\tiny [0.2,0.2]} \\
\cline{1-1} \cline{3-4} \cline{6-7}
H+P & & 0.3 {\tiny [0.2,0.3]} & 0.3 {\tiny [0.2,0.3]} & & 0.07 {\tiny [0.07,0.08]} & 0.2 {\tiny [0.2,0.2]} \\
\cline{1-1} \cline{3-4} \cline{6-7}
\multicolumn{2}{c}{}&\multicolumn{5}{c}{\textbf{SI-SDR} $\uparrow$} \\
\cline{1-1} \cline{3-4} \cline{6-7}
H   & & \cellcolor{gray!30}37.47 {\tiny [36.76,38.16]} & 35.31 {\tiny [34.53,36.10]} & & \cellcolor{gray!30}43.85 {\tiny [43.00,44.69]}& 28.87 {\tiny [28.67,29.06]} \\
P   & & 37.21 {\tiny [36.51,37.89]} & \cellcolor{gray!30}35.50 {\tiny [34.72,36.28]} & & 43.83 {\tiny [43.05,44.61]} & \cellcolor{gray!30}29.14 {\tiny [28.95,29.33]} \\
\cline{1-1} \cline{3-4} \cline{6-7}
H+P & & 37.59 {\tiny [36.88,38.30]} & 35.90 {\tiny [35.09,36.69]} & & 44.18 {\tiny [43.38,45.00]} & 29.17 {\tiny [28.98,29.36]} \\
\cline{1-1} \cline{3-4} \cline{6-7}
\multicolumn{2}{c}{}&\multicolumn{5}{c}{\textbf{ViSQOL} $\uparrow$} \\
\cline{1-1} \cline{3-4} \cline{6-7}
H   & & 3.04 {\tiny [2.99,3.09]} & 3.07 {\tiny [3.02,3.13]} & & \cellcolor{gray!30}4.14 {\tiny [4.09,4.18]} & 3.65 {\tiny [3.62,3.67]} \\
P   & & \cellcolor{gray!30}3.09 {\tiny [3.04,3.14]} & \cellcolor{gray!30}3.08 {\tiny [3.03,3.14]} & & \cellcolor{gray!30}4.14 {\tiny [4.10,4.18]} & \cellcolor{gray!30}3.76 {\tiny [3.73,3.78]} \\
\cline{1-1} \cline{3-4} \cline{6-7}
H+P & & 3.21 {\tiny [3.15,3.26]} & 3.18 {\tiny [3.12,3.23]} & & 4.20 {\tiny [4.16,4.24]} & 3.80 {\tiny [3.77,3.82]} \\
\cline{1-1} \cline{3-4} \cline{6-7}
\end{tabular}
}

\vspace{10pt}

\scalebox{0.8}{
\begin{tabular}{|c|c|c|c|c|c|c|}
\multicolumn{2}{c}{} & \multicolumn{1}{c}{\textit{Percussive}} & \multicolumn{1}{c}{}& \multicolumn{1}{c}{\textit{Environ. sounds}} &\multicolumn{1}{c}{}& \multicolumn{1}{c}{\textit{Speech}} \\
\cline{3-3} \cline{5-5} \cline{7-7}
\multicolumn{2}{c|}{} & ENST-drums &\multicolumn{1}{c|}{}& ESC-50 &\multicolumn{1}{c|}{}& VCTK \\
\cline{3-3} \cline{5-5} \cline{7-7}
\multicolumn{2}{c}{}&\multicolumn{5}{c}{\textbf{Mel} $\downarrow$} \\
\cline{1-1} \cline{3-3} \cline{5-5} \cline{7-7}
H   & & 0.45 {\tiny [0.44,0.46]} & & 0.48 {\tiny [0.46,0.50]} & & 0.74 {\tiny [0.73,0.75]} \\
P   & & \cellcolor{gray!30}0.42 {\tiny [0.41,0.43]} & & \cellcolor{gray!30}0.47 {\tiny [0.44,0.49]} & & \cellcolor{gray!30}0.67 {\tiny [0.66,0.67]} \\
\cline{1-1} \cline{3-3} \cline{5-5} \cline{7-7}
H+P & & 0.41 {\tiny [0.40,0.42]} & & 0.46 {\tiny [0.43,0.48]} & & 0.70 {\tiny [0.69,0.70]} \\
\cline{1-1} \cline{3-3} \cline{5-5} \cline{7-7}
\multicolumn{2}{c}{}&\multicolumn{5}{c}{\textbf{STFT} $\downarrow$} \\
\cline{1-1} \cline{3-3} \cline{5-5} \cline{7-7}
H   & & 1.74 {\tiny [1.71,1.77]} & & \cellcolor{gray!30}2.10 {\tiny [2.04,2.15]} & & 2.91 {\tiny [2.89,2.93]} \\
P   & & \cellcolor{gray!30}1.66 {\tiny [1.64,1.69]} & & \cellcolor{gray!30}2.10 {\tiny [2.04,2.15]} & & \cellcolor{gray!30}2.73 {\tiny [2.71,2.75]} \\
\cline{1-1} \cline{3-3} \cline{5-5} \cline{7-7}
H+P & & 1.63 {\tiny [1.60,1.66]} & & 2.07 {\tiny [2.01,2.12]} & & 2.79 {\tiny [2.77,2.81]} \\
\cline{1-1} \cline{3-3} \cline{5-5} \cline{7-7}
\multicolumn{2}{c}{}&\multicolumn{5}{c}{\textbf{Waveform} $\boldsymbol{\downarrow}$ \textbf{(}$\boldsymbol{10^{-2}}$\textbf{)}} \\
\cline{1-1} \cline{3-3} \cline{5-5} \cline{7-7}
H   & & \cellcolor{gray!30}0.7 {\tiny [0.6,0.7]} & & \cellcolor{gray!30}1.0 {\tiny [0.9,1.1]} & & \cellcolor{gray!30}0.5 {\tiny [0.4,0.5]}\\
P   & & \cellcolor{gray!30}0.7 {\tiny [0.6,0.7]} & & \cellcolor{gray!30}1.0 {\tiny [0.9,1.1]} & & \cellcolor{gray!30}0.5 {\tiny [0.5,0.5]} \\
\cline{1-1} \cline{3-3} \cline{5-5} \cline{7-7}
H+P & & 0.7 {\tiny [0.6,0.7]} & & 1.0 {\tiny [0.9,1.1]} & & 0.5 {\tiny [0.5,0.5]} \\
\cline{1-1} \cline{3-3} \cline{5-5} \cline{7-7}
\multicolumn{2}{c}{}&\multicolumn{5}{c}{\textbf{SI-SDR} $\uparrow$} \\
\cline{1-1} \cline{3-3} \cline{5-5} \cline{7-7}
H   & & \cellcolor{gray!30}18.91 {\tiny [18.09,19.74]} & & \cellcolor{gray!30}19.89 {\tiny [18.97,20.83]} & & \cellcolor{gray!30}21.29 {\tiny [20.94,21.64]} \\
P   & & 18.53 {\tiny [17.72,19.35]} & & 19.52 {\tiny [18.57,20.48]} & & 20.35 {\tiny [20.00,20.70]} \\
\cline{1-1} \cline{3-3} \cline{5-5} \cline{7-7}
H+P & & 18.82 {\tiny [17.97,19.64]} & & 19.86 {\tiny [18.93,20.82]} & & 20.86 {\tiny [20.50,21.21]} \\
\cline{1-1} \cline{3-3} \cline{5-5} \cline{7-7}
\multicolumn{2}{c}{}&\multicolumn{5}{c}{\textbf{ViSQOL} $\uparrow$} \\
\cline{1-1} \cline{3-3} \cline{5-5} \cline{7-7}
H   & & 3.26 {\tiny [3.21,3.31]} & & 2.95 {\tiny [2.90,2.99]} & & 2.45 {\tiny [2.43,2.47]} \\
P   & & \cellcolor{gray!30}3.33 {\tiny [3.28,3.38]} & & \cellcolor{gray!30}3.04 {\tiny [2.99,3.08]} & & \cellcolor{gray!30}2.50 {\tiny [2.48,2.52]} \\
\cline{1-1} \cline{3-3} \cline{5-5} \cline{7-7}
H+P & & 3.40 {\tiny [3.34,3.45]} & & 3.14 {\tiny [3.09,3.18]} & & 2.50 {\tiny [2.48,2.52]} \\
\cline{1-1} \cline{3-3} \cline{5-5} \cline{7-7}
\end{tabular}
}

\label{tab-sections}
\end{table}

\subsection{Harmonic-Percussive analysis}

In the experiment illustrated by Table \ref{tab-sections}, we separately used $\{(\Tilde{H}_n^{48;1}), (\Tilde{H}_n^{48;2}), (\Tilde{H}_n^{48;3})\}$ the estimated harmonic tokens, and $\{(\Tilde{P}_n^{48;1}), (\Tilde{P}_n^{48;2}), (\Tilde{P}_n^{48;3})\}$ the estimated percussive tokens, to evaluate the reconstruction associated with each group of tokens, and each combination of these groups. We applied this differentiated procedure of estimation to a percussive dataset (ENST-drums), two purely harmonic datasets (Monophonic and Polyphonic), two general music datasets (Medley-solos-DB and OrchideaSOL), a speech dataset (VCTK) and an environmental sounds dataset (ESC-50). The reconstruction metrics we obtained show that the harmonic section of the estimated tokens better reconstruct a harmonic signal, while the percussive section of the estimated tokens better reconstruct a percussive signal. This underlines the interest of our architecture.

\section{Limitations}

\noindent Our approach is based on two models: HP-codec, a neural audio codec and a language model (which together form HP-codecX). This constitutes the main drawback of this work, as both models must be trained jointly in order to have a fully working process. A promising direction for future work would be to relax this architectural coupling at the codec level, building further on the direction explored in \textbf{Exp 5} of our ablation study, and drawing inspiration from recent approaches such as \cite{li2024masksrmaskedlanguagemodel} and \cite{yang2024genhancer}. These works employ an off-the-shelf codec to produce incomplete discrete token sequences, and then train a generative language model to recover the corresponding clean representations. Although these techniques have so far been explored only in speech domains, they offer a promising direction for future research on the task considered here. \\
In contrast to various bandwidth-extension systems such as \cite{li2025apollo} and \cite{liu2024audiosr}, our model does not support variable input sampling rates and is currently limited to mapping 16~kHz inputs to 48~kHz outputs. This constraint arises primarily from our reliance on discrete audio codecs, which themselves typically operate at fixed sampling rates. Nevertheless, the 16~kHz-48~kHz setting already yields a substantial and practically meaningful improvement in spectral coverage, demonstrating the viability of the proposed framework. Moreover, the relatively low training cost of each model instance makes it feasible to train separate variants for additional sampling rates when needed.\\

\section{Conclusion}

\noindent We introduce HP-Codec, a multi-branch neural audio codec that produces a latent representation disentangled across frequency bands. This disentanglement is further enhanced through a Harmonic–Percussive decomposition, which strengthens inter-band coupling and facilitates prediction of high-frequency representations from their low-frequency counterparts. In this way, we restructure the codec architecture to naturally support the downstream task of bandwidth extension.
Building upon this design, we propose HP-CodecX, a bandwidth extension model that integrates HP-Codec with a transformer-based audio language model. The transformer mirrors the codec’s architecture, enabling effective modeling of cross-band dependencies. Empirical results across multiple datasets demonstrate that HP-CodecX achieves state-of-the-art performance on both objective and subjective metrics, yielding more accurate high-frequency reconstruction than existing baselines.

\section*{Acknowledgments}
\noindent This work was funded by the European Union (ERC, HI-Audio, 101052978). Views and opinions expressed are however those of the authors only and do not necessarily reflect those of the European Union or the European Research Council. Neither the European Union nor the granting authority can be held responsible for them.

\bibliography{refs_long,TASLP2026}

\vfill

\end{document}